\documentclass[journal=ancac3,manuscript=article,layout=twocolumn]{achemso}
\setkeys{acs}{maxauthors=20, etalmode=truncate}
\setkeys{acs}{articletitle=true}
\SectionsOn

\usepackage[version=3]{mhchem} 
\usepackage[T1]{fontenc}       
\usepackage{graphicx, color,amssymb}
\usepackage{hyperref}

\newcommand{\ie}{\emph{i.\,e.}}

\newcommand{\ea}{\emph{et al.}}

\newcommand{\si}{Supporting Information}

\usepackage{eso-pic}

\newcommand{\publisherstatement}{%
  \AddToShipoutPictureFG*{%
    \AtPageUpperLeft{%
      \raisebox{-12mm}{
        \hspace*{15mm}
        \parbox{0.88\paperwidth}{%
          \small
          \fbox{%
            \parbox{0.86\paperwidth}{%
              This document is the Accepted Manuscript version of a Published Article that appeared in final form in \emph{ACS Nano}, copyright \textcopyright\ 2024 American Chemical Society. The final published version is available at \href{https://doi.org/10.1021/acsnano.4c05123}{doi:10.1021/acsnano.4c05123}.

            }%
          }%
        }%
      }%
    }%
  }%
}

\newcommand{\ude}{\affiliation{Faculty of Physics and Center for Nanointegration Duisburg-Essen (CENIDE), University of Duisburg-Essen; Duisburg, 47057, Germany.}}
\newcommand{\cau}{\affiliation{Institute for Inorganic Chemistry, Christian-Albrechts-University; Kiel, 24098, Germany.}}
\newcommand{\xfel}{\affiliation{European XFEL; Schenefeld, 22869, Germany.}}
\newcommand{\iqmt}{\affiliation{Institute for Quantum Materials and Technologies (IQMT), Karlsruhe Institute for Technology (KIT); Eggstein-Leopoldshafen, 76344, Germany.}}
\newcommand{\cesq}{\alsoaffiliation{Centre Européen de Sciences Quantiques (CESQ), Institut de Science et d’Ingénierie Supramoléculaires (ISIS); Strasbourg Cedex, 67083, France.}}
\newcommand{\wildau}{\affiliation{Faculty of Engineering and Natural Sciences, Technical University of Applied Science Wildau; \\Wildau, 15745, Germany.}}

\author{Lea Kämmerer} \ude \altaffiliation{Contributed equally to this work}
\author{Gérald Kämmerer} \ude \altaffiliation{Contributed equally to this work}
\author{Manuel Gruber} \email{manuel.gruber@uni-due.de} \ude 
\author{Jan Grunwald} \cau
\author{Tobias Lojewski} \ude
\author{Laurent Mercadier} \xfel
\author{Loïc Le Guyader} \xfel
\author{Robert Carley} \xfel
\author{Cammille Carinan} \xfel
\author{Natalia Gerasimova} \xfel
\author{David Hickin} \xfel
\author{Benjamin E. Van Kuiken} \xfel
\author{Giuseppe Mercurio} \xfel
\author{Martin Teichmann} \xfel
\author{Senthil Kumar Kuppusamy} \iqmt 
\author{Andreas Scherz} \xfel
\author{Mario Ruben} \iqmt \cesq
\author{Klaus Sokolowski-Tinten} \ude
\author{Andrea Eschenlohr} \ude
\author{Katharina Ollefs} \ude
\author{Carolin Schmitz-Antoniak} \wildau
\author{Felix Tuczek} \cau
\author{Peter Kratzer} \ude
\author{\\Uwe Bovensiepen} \ude
\author{Heiko Wende} \ude

\title[\textsf{achemso}]{Femtosecond spin-state switching dynamics of Fe(II) complexes condensed in thin films}

\begin{document}
\publisherstatement

\footnotesize
\setlength{\fboxrule}{0 pt}
\begin{tocentry}
\begin{center}
\includegraphics{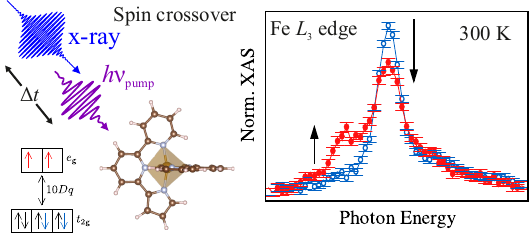}
\end{center}
\end{tocentry}

\subsection*{Abstract} 
The tailoring of spin-crossover films has made significant progress over the last decade, mostly motivated by the prospect in technological applications.
In contrast to spin-crossover complexes in solution, the investigation of the ultrafast switching in spin-crossover films has remained scarce. 
Combining the progress in molecule synthesis and film growth with the opportunities at x-ray free-electron lasers, we study the photo-induced spin-state switching dynamics of a molecular film at room temperature.
The sub-picosecond switching from the $S=0$ low-spin ground state to the $S=2$ high-spin state is monitored by analyzing the transient evolution of the Fe $L_3$ x-ray absorption edge fine structure, \ie\ element-specifically at the reaction center of the Fe(II) complex.
Our measurements show the involvement of an intermediate state in the switching.
At large excitation fluences, the fraction of high-spin molecules saturates at $\approx 50$\,\%, which is likely due to molecule-molecule interaction within the film.

\textbf{Keywords:} spin crossover; ultrafast switching; molecular film; x-ray free electron laser; time-resolved x-ray absorption spectroscopy


Spin-crossover (SCO) complexes may be switched between a low- and a high-spin state in response to various stimuli such as temperature, light, and pressure \cite{gutlich_spin_2000,halcrow_spin-crossover_2013}.
The spin transition is accompanied with a change of geometry, electronic and optical properties of the compounds. 
The preparation of SCO thin films, combined with the multitude of stimuli and read out, provides a rich variety of potential applications, such as sensors, displays, actuators, and data storage \cite{kahn_spin_1992,shepherd_spin_2012,molnar_emerging_2014,molnar_spin_2017,kumar_emerging_2017}.

The quality of SCO films, in terms of purity and homogeneity, has significantly increased over the last decade by employing sublimation in (ultrahigh) vacuum rather than wet-chemistry deposition methods \cite{shi_study_2009,naggert_first_2011,bernien_spin_2012,palamarciuc_spin_2012, warner_temperature-_2013,davesne_hysteresis_2015,naggert_vacuum-evaporable_2015,iasco_disentangling_2017,shalabaeva_vacuum_2017,ridier_hysteresis_2017, zhang_locking_2017,poggini_room_2018, mallah_surfaces_2018,atzori_thermal_2018, kipgen_evolution_2018,rohlf_light-induced_2018, kumar_sublimable_2019,rohlf_influence_2019,poggini_surface_2019,ossinger_spin_2020, kipgen_spincrossover_2021,kelai_thermal_2021, rohlf_probing_2021, grunwald_defying_2023,chen_observation_2023}.
The major difficulty for preparing films via sublimation has been the synthesis and identification of robust SCO compounds withstanding the sublimation temperature and the adsorption on the substrate \cite{ossinger_vacuum-evaporable_2017,knaak_fragmentation_2019,gruber_spin-crossover_2020,johannsen_spin-crossover_2023}, issues which have to a large extent been successfully tackled in the last years \cite{bairagi_molecular-scale_2016,kipgen_soft-x-ray-induced_2017,johannsen_electron-induced_2021,penicaud_substrate-dependent_2023}.

\begin{figure*}[tb]
	\includegraphics[scale=1]{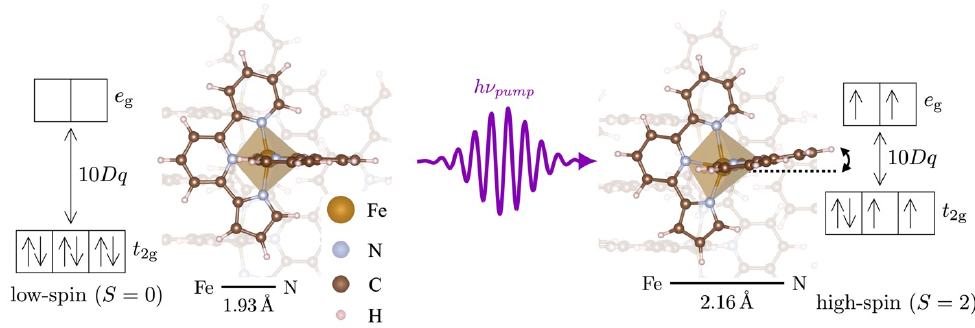}
	\caption{\footnotesize \textbf{Low- and high-spin configurations in spin-crossover molecules.}
The calculated geometries of an individual [Fe(pypypyr)$_2$] molecule in the low-spin ($S = 0$, left) and high-spin ($S = 2$, right) state.
Concomitantly with the low- to high-spin transition, the Fe-N bond length changes by approximately 10\,\% as indicated by the corresponding average bond lengths inferred from density functional theory calculations.
The corresponding ligand twisting in the high-spin state is indicated on the right by the dashed line and the bended arrow.
The excitation of the molecule by an external stimulus, such as a pump laser pulse at photon energy $h \nu_\text{pump}$, can trigger this transition.
The splitting of $3d$ states of the Fe$^{2+}$ ion into $t_{2g}$ and $e_g$ in an octahedral ligand field is represented by the 10Dq value as indicated at the left and right side for $S = 0$ and $S = 2$, respectively.
In the condensed form, the molecules are surrounded by peers, as illustrated with molecules in light colors in the background, which can further modify the switching properties.
	\label{fig:molecule}}
\end{figure*}

Besides their importance for potential applications, SCO thin films also exhibit a richness of physical properties.
The detailed local environment of the SCO complexes as well as intermolecular interactions influence the SCO properties of the films \cite{bairagi_molecular-scale_2016,kipgen_evolution_2018,rohlf_influence_2019,fourmental_importance_2019,ossinger_effect_2020,johannsen_electron-induced_2021,johannsen_spin_2022}.
Intermolecular interaction in the condensed phase is of particular interest to induce cooperativity, which leads to a sharp thermal spin-state transition of the film.
This transition may be thermally shifted and incomplete, \ie\ with a fraction of non-switched molecules, because of the different thermodynamic properties of the molecules at the interfaces and because of steric repulsion between molecules \cite{felix_enhanced_2013,gruber_spin_2017,kelai_thermal_2021}.
Overall the organization of the molecules in thin films give further prospects in engineering the SCO properties of a given compound.

X-ray absorption spectroscopy (XAS) at the Fe $L_{3}$ edges is particularly adapted to investigate the spin-state transition of SCO thin films\cite{cartier_dit_moulin_spin_1992,davesne_first_2013,kuch_controlling_2017,gruber_spin-crossover_2020}. 
This edge (transition from $2p_{3/2}$ to $3d$ states) is sensitive to the electronic population of the Fe $d$ orbitals, which is changing upon spin crossover.
As further detailed below, the low-spin (LS) and high-spin (HS) states are characterized by different spectroscopic fine structures, which allows quantitative determination of the HS fraction.

Despite the recent development in SCO thin films, the investigation of the switching dynamics upon optical excitation has remained scarce and mostly limited to slow evolution of the fraction of high-spin molecules at low temperatures \cite{kipgen_soft-x-ray-induced_2017,rohlf_light-induced_2018,zhang_anomalous_2020,grunwald_defying_2023}, although the ultrafast switching dynamics has been studied in detail for individual Fe(II) complexes in solution \cite{gawelda_ultrafast_2007,smeigh_femtosecond_2008,lorenc_successive_2009, wolf_sub-picosecond_2008,bressler_femtosecond_2009,zhang_tracking_2014,azzolina_single_2019, lemke_coherent_2017,huse_femtosecond_2011,huse_photo-induced_2010}.
Few SCO dynamics studies on powders and single crystals are also available \cite{freyer_ultrafast_2013,field_spectral_2016}.
Ridier \ea\ investigated SCO films with thicknesses on the order of 100\,nm using pump-probe optical methods \cite{ridier_finite_2019}.
They observed a sub-picosecond spin-state switching followed by a slow dynamics on the order of 10\,ns attributed to the heating of the film.
Zhang \ea\ studied films of [Fe(phen)$_3$]$^{2+}$ of similar thicknesses and observed an evolution of the Fe $M$ edge spectra with pump-probe delay, interpreted in terms of a spin-state transition through a metal-centered triplet intermediate state \cite{zhang_tracking_2019}.
The ultrafast switching of much thinner charge-neutral SCO films, requiring a sensitive method with large spectral changes such as XAS at the Fe $L_3$ edge, has not yet been reported.

Intermolecular interactions in the condensed phase likely influence the dynamics at different time scales.
At very short time scales ($\lesssim 1$\,ps) the spin state switching dynamics presumably depends on the local surrounding of a given molecule which in turn is influenced by the spin state of the neighboring molecules.
Such local stress may extend and propagate in the materials, which would modify the dynamics over longer time scales.
Ultrafast investigations should provide insights about the impact of the local environment on the switching dynamics.

Here, we reveal the room-temperature ultrafast switching of $\approx 10$\,nm ultrathin film of [Fe(pypypyr)$_2$] (pypypyr = bipyridyl pyrrolide) sublimated on Si$_3$N$_4$.
The transient evolution of the absorption fine structure at the Fe $L_3$ edge from the LS to HS state is triggered by optical pumping and analyzed with 80 fs time resolution.
The results showcase the transient population of an intermediate state in the switching dynamics.
By increasing the density of optically excited molecules in the material, we identify a limit of $\approx 50$\,\% of switched molecules, which is rationalized primarily by the large molecular distortion that leads to local stress.
For switching a large fraction of molecules, such stress must be relaxed locally, presumably by unswitched molecules adjacent to switched ones.

\begin{figure*}[hbt]
	\includegraphics[width=0.6\textwidth]{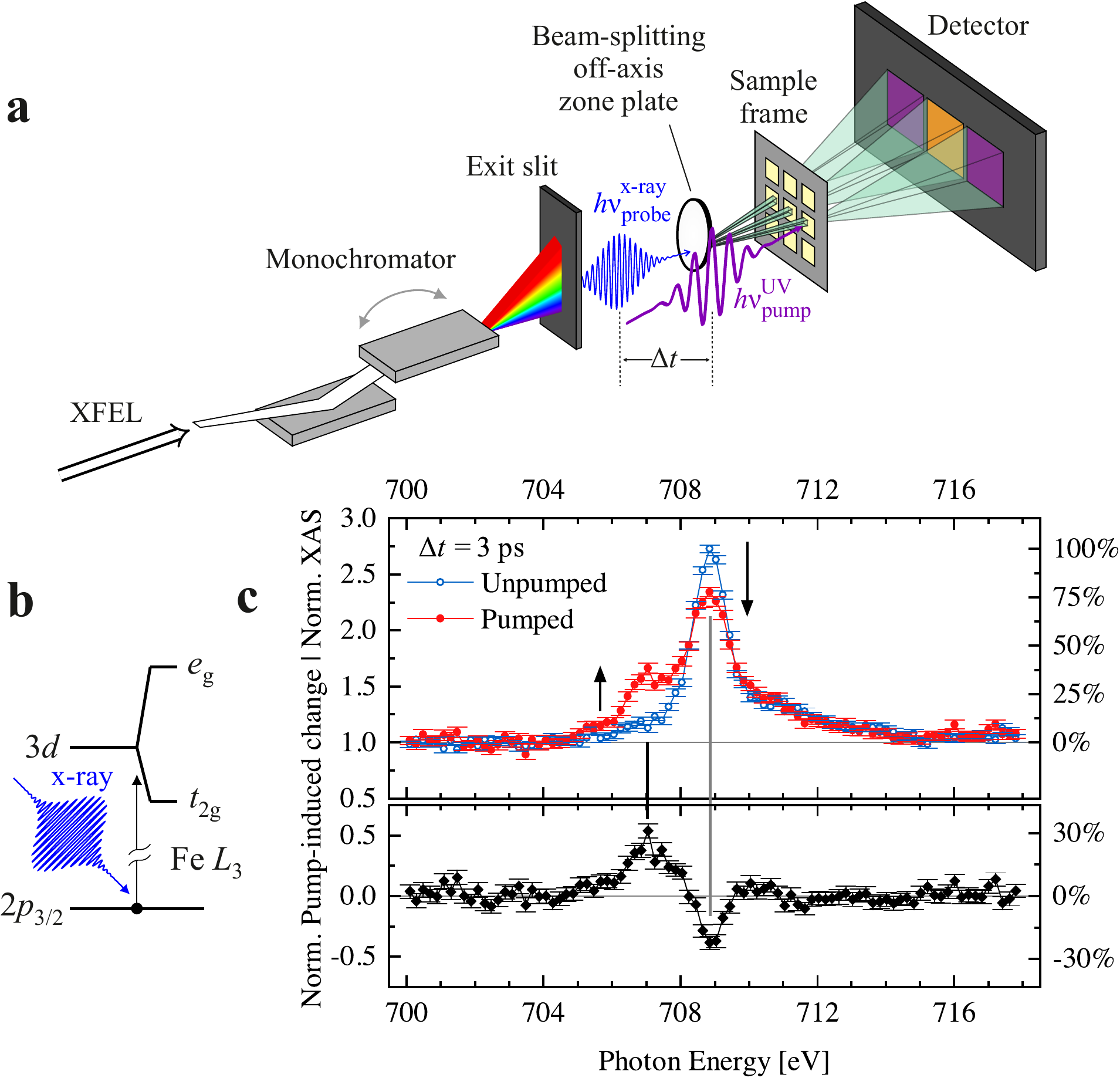}
	\caption{\footnotesize \textbf{Time-resolved x-ray absorption spectroscopy at the Fe $L_3$ edge}.
(a) Schematic of the experimental setup of the Spectroscopy and Coherent Scattering Instrument at the European XFEL.
(b) Schematic x-ray absorption process for the measured fine structure at the Fe $L_3$ absorption edge.
(c) An x-ray absorption spectrum of the molecular film measured at room temperature at the unpumped window (open circles, blue) is depicted together with a pumped spectrum (filled circles, red) at time delay $\Delta t = 3$\,ps, which was recorded at a laser pumped membrane window.
Both spectra are corrected with the simultaneously measured $I_0$ on a reference Si$_3$N$_4$ window (\si).
The difference of the pumped and the unpumped spectrum represents the pump-induced change (bottom panel).
The right axis is given relative to the while line maximum of the unpumped spectrum.
The two features in the spectrum at photon energies of 707.1\,eV and 708.9\,eV, see vertical lines, show changes in intensities that are highlighted by arrows.
This spectral modification is represented by the pump-induced change and corresponds to the spin-state switching from $S = 0$ to $S = 2$, induced by the pump laser at an incident fluence of $10$ \,mJ/cm$^2$.
	\label{fig:scs}}
\end{figure*}

\subsection*{Results/Discussion}

Fig.~\ref{fig:molecule} shows a sketch of the [Fe(pypypyr)$_2$] complex in the LS and in the HS configurations.
In the LS state, the $t_{2g}$ orbitals are fully filled, forming a singlet LS state with $S=0$.
In the HS state, the ligand field strength is reduced due to the increased average Fe-N distance.
The $t_{2g}$ and $e_g$ orbitals are partially occupied forming a quintet ($S=2$).
The characterization of [Fe(pypypyr)$_2$] films upon sublimation as well as the possibility to optically switch to a quintet HS state are discussed in Ref.~\citenum{grunwald_defying_2023}.
The thin film of [Fe(pypypyr)$_2$] complex is in the LS state at room temperature, with presumably a fast HS to LS relaxation time.
These properties make such films, a priori, suitable for stroboscopic investigations.
Such films are deposited on Si$_3$N$_4$ membranes, allowing measurements in a transmission mode.

We used the Spectroscopy and Coherent Scattering instrument (SCS) at the European XFEL to probe the ultrafast dynamics of an ultrathin film of [Fe(pypypyr)$_2$] deposited on Si$_3$N$_4$ by sublimation (see \si).
The instrument provides a shot-to-shot normalization scheme with high signal-to-noise ratio in the x-ray absorption spectra (Fig.~\ref{fig:scs}a) \cite{le_guyader_photon-shot-noise-limited_2023,lojewski_interplay_2023,porro_minisdd-based_2021}.
The SASE3 undulator system generates x-ray pulses, which are monochromatized by a variable line-spacing grating in combination with an exit slit.
The x-rays then pass through beam-splitting off-axis zone plate optics which split the x-ray beam into three beams of equal intensity.
These beams are transmitted through the sample structure consisting of one bare Si$_3$Ni$_4$ membrane and two molecular films on such membranes, one of them is optically excited by the pump laser.
The three beams are detected on an imaging detector\cite{porro_minisdd-based_2021} to distinguish the three signals.
This setup allows for shot-to-shot normalization, which is essential for a fluctuating light source as used here, see Ref.~\citenum{le_guyader_photon-shot-noise-limited_2023} for details.

\begin{figure*}[hbt]
	\includegraphics[width=0.9\textwidth]{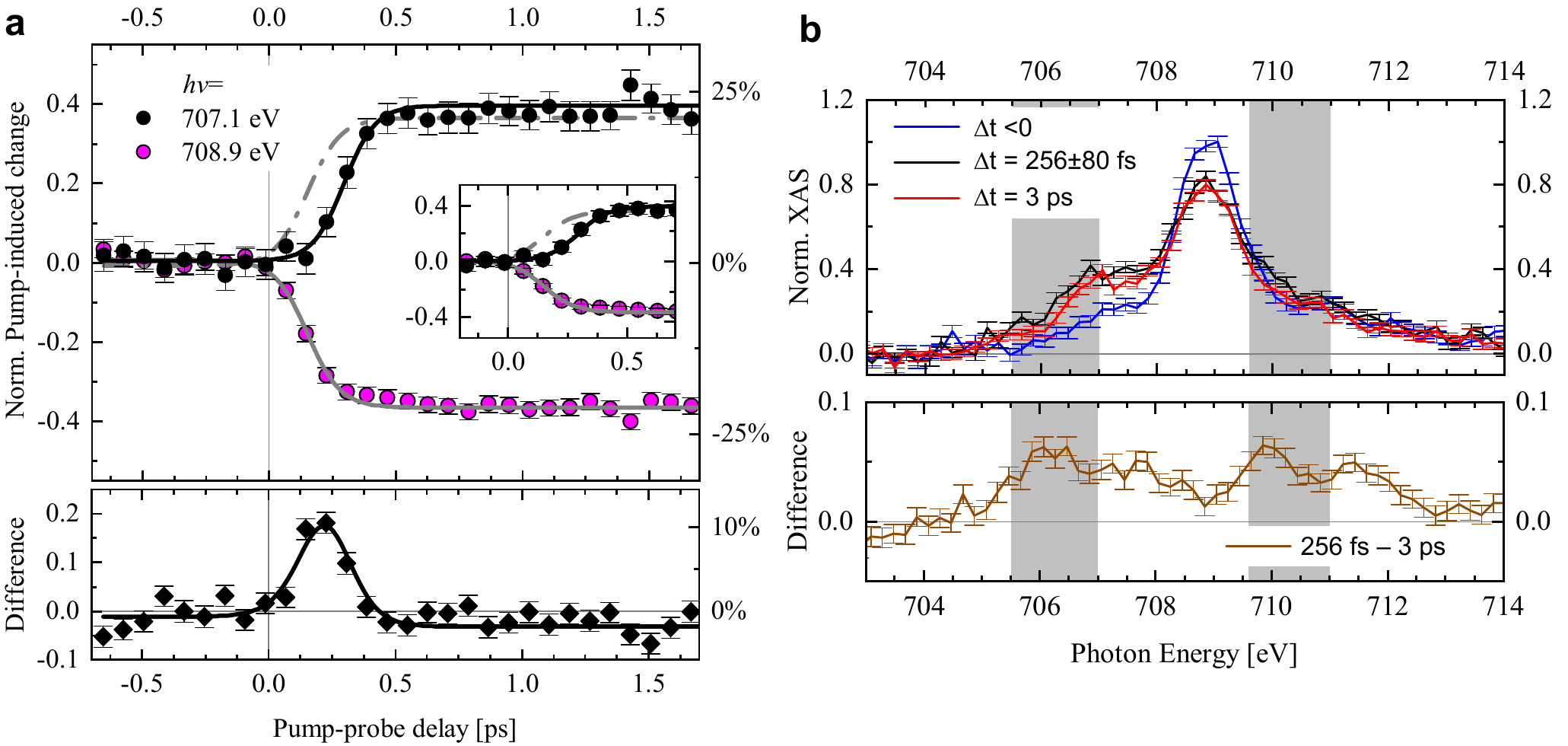}
	\caption{\footnotesize \textbf{Intermediate state analysis of the spin-crossover process.}
(a) Top panel: Normalized pump-induced change at 707.1\,eV (black symbols) and 708.9\,eV (pink symbols) photon energy as a function of pump-probe delay $\Delta t$.
The lines guide the eye.
The gray dashed line is the absolute value of the guide to the eye of the negative transient for 708.9\,eV. The right ordinate indicates the ratio to the maximum in the unpumped spectrum, see Fig.~\ref{fig:scs}c.
Inset: The data of the main panel are presented for a reduced time delay range.
Bottom panel: The difference of the two transient absorption traces shown in the top panel are depicted.
The data are reconstructed out of 26 time-delay sweeps, for which the time zero reference is adjusted as discussed in the SI.
The timing of the two transients on the time delay has been defined to a separate reference measurement and does not take drifts into account.
(b) X-ray absorption spectra at three different time delays $\Delta t$: before pumping (blue), at 256\,fs (black), and at 3\,ps (red).
In the region of the gray bar the spectrum at 256\,fs exhibits larger absorption than the one at 3 ps.
The spectrum $\Delta t = 256\pm80$\,fs is an average of three transient ($\Delta t = 191$, 238, and 338\,fs).
The concomitantly measured spectra on unpumped windows (in the LS state) were normalized to their peak intensity at a photon energy of 708.9\,eV. 
The same normalization factors were used to scale the corresponding pumped spectra.
The lower panel shows the difference between the $\Delta t = 256$\,fs and the $\Delta t = 3$\,ps spectra (filtered with a 5-point moving average).
	\label{fig:timeEvolution}}
\end{figure*}

The x-ray absorption spectra were measured in transmission geometry in a pump-probe configuration by sweeping the photon energy of ultrashort x-ray pulses [full width at half maximum (FWHM) of 30--60\,fs] synchronized with an ultrashort pump pulse at 3.1\,eV photon energy and a FWHM pulse duration of 50\,fs.
For further technical details, see \si\ and Ref.~\citenum{le_guyader_photon-shot-noise-limited_2023}.
The chosen photon energy of the optical pump pulse induces a resonant excitation from the low-spin ground state to a metal-to-ligand charge transfer (MLCT) state, in turn relaxing to the high-spin state \cite{hauser_light-induced_2004,chergui_photoinduced_2017}.
The x-ray pulse probes the transient state after a pump-probe time delay $\Delta t$, see Fig.~\ref{fig:scs}a.
Resonant absorption at the Fe $L_3$ absorption edge probes the unoccupied electronic states of the $t_{2g}$ and $e_g$ orbitals (Fig.~\ref{fig:scs}b).
In the absence of pumping, the x-ray absorption spectrum exhibits a main peak at 708.9\,eV indicating a low-spin state of the molecules at room temperature.
A pumped spectrum at $\Delta t=3$\,ps exhibits an increase and decrease in x-ray absorption at 707.1\,eV and 708.9\,eV, respectively, which is shown in Fig.~\ref{fig:scs}c in comparison to a spectrum obtained without pumping.
The difference of the pumped and unpumped spectra is shown in the lower panel of Fig.~\ref{fig:scs}c as the pump-induced change which is clearly identified and reaches 30\,\%.
This value represents a change of population from the low to the high spin state \cite{grunwald_defying_2023}.
We note that the spectrum acquired 44\,$\mu s$ after the pump pulse, which is possible via a shot-to-shot analysis of the absorption, is essentially identical to the spectrum prior to excitation (Figure S6).
The relaxation from the high- to the low-spin state therefore occurs in less than 44\,$\mu s$ at room temperature.
In addition, the full reversibility of the switching is evidence for a negligible radiation-induced fragmentation of the molecules.

Fig.~\ref{fig:timeEvolution}a shows the evolution of the x-ray absorption at 707.1 and 708.9 eV as a function of pump-probe time delay.
While the absorption changes by $\approx25$\,\% for both delay sweeps, the saturation is reached faster within 0.3\,ps at 708.9 eV than at 707.1 eV, where reaching saturation takes 0.5 ps.
The difference between the two transients is significant for 0<$\Delta t$<0.5 ps as depicted in the bottom panel of Fig. \ref{fig:timeEvolution}a, which indicates the transient population of an intermediate state.
We emphasize that the transition from the low- to the high-spin configuration involves an electronically excited state in combination with a change of nuclear coordinates.
We acquired transient x-ray absorption spectra at the early instants of the dynamics on sub-picosecond timescales shown in Fig.~\ref{fig:timeEvolution}b and in the \si\ along with spectra at time delays $\Delta t < 0$ and $\Delta t = 3$\,ps.
The spectrum at $\Delta t = 3$\,ps has an increased intensity at 707.1\,eV and a lower intensity at 708.9\,eV compared to the one prior to the pumping ($\Delta t < 0$).
The transient spectrum at $\Delta t \approx 256 \pm 80$\,fs is similar the one acquired at later time delays, indicating that a large fraction of the molecules already has switched to the HS state (in addition to approximately half of the molecules not participating in the switching and remaining in the LS state).
There are, however, additional features compared to spectra recorded before pumping and at $\Delta t=3$\,ps.
A small increase in x-ray absorption is observed pre-edge at 706.3\,eV and, although weaker, post-edge at 710\,eV.
These small absorption increases are systematic and larger than the respective uncertainty bars (see \si).

Simulations of x-ray absorption spectra using atomic multiplet and ligand-field theory are provided in the \si, Figure~S3.
Comparisons of these simulated spectra with the 256-fs transient spectrum suggest that the resolved intermediate state is a metal-centered $^3T_2$ triplet state, in line with previous reports \cite{zhang_manipulating_2017,kjaer_finding_2019,zhang_tracking_2019,alias-rodriguez_ultrafast_2023}.
It should be noted that the involvement of a ligand-centered $^3$MLCT state, prior to the $^3T_2
$ population, cannot be excluded.
The absence of other structure in our data indicate that the lifetime of the $^3$MLCT state, if populated, is even shorter than that of the $^3T_2$ state.
The different time evolution of the two x-ray absorption features shown in Fig.~\ref{fig:timeEvolution}a can be well reproduced using a rate equation model assuming a single intermediate state with a finite relaxation time, see \si.

\begin{figure}[hbt]
	\includegraphics[width=\columnwidth]{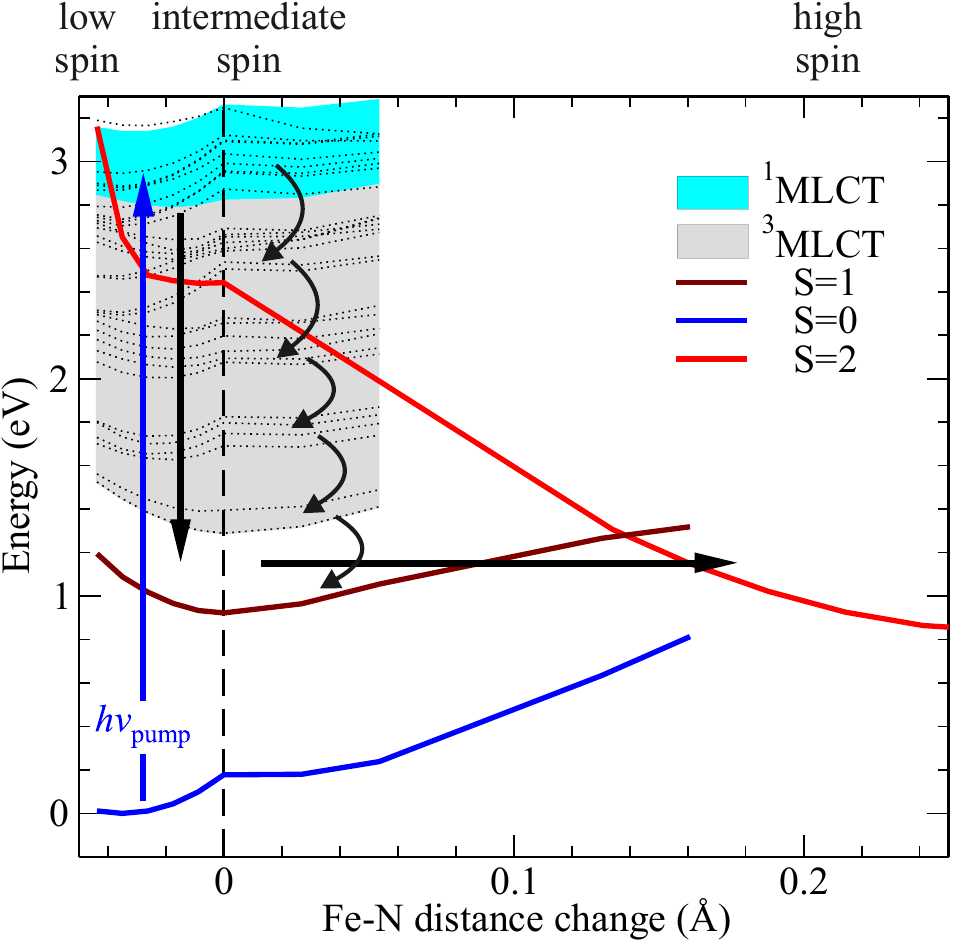}
	\caption{\footnotesize \textbf{Calculated potential energy surfaces.}
Total energies of the metal-centered $S = 0$, $S = 1$, and $S = 2$ (thick lines) states, as well as $^1$MLCT singlet (turquoise area), and $^3$MLCT (gray area) as a function of the average change in Fe-N bond length as inferred from time-dependent DFT.
As a starting point, we consider a photo-excitation from the $S=0$ to the manifold of $^1$MLCT states.
The arrows illustrate hypothetical relaxation pathways.
The Fe-N distance change varies along two coordinates to represent the low-, intermediate-, and high-spin states.
	\label{fig:energySurface}}
\end{figure}

The photoinduced dynamics of complexes \textit{in solutions} have been studied using pump-probe experiments employing optical \cite{gawelda_ultrafast_2007,smeigh_femtosecond_2008,lorenc_successive_2009, wolf_sub-picosecond_2008}, hard \cite{bressler_femtosecond_2009,zhang_tracking_2014,azzolina_single_2019, lemke_coherent_2017}, and soft x-ray \cite{huse_femtosecond_2011,huse_photo-induced_2010} probes along with ab initio calculations \cite{degraaf_study_2010,sousa_ultrafast_2013} over the last decades.
These works lead to the broadly accepted microscopic mechanism upon optical excitation, which proceeds on femtosecond timescales.
It involves the population of a metal-to-ligand-charge-transfer (MLCT) singlet state starting from the low-spin ground state, followed by an ultrafast relaxation along the multidimensional Fe-N nuclear coordinate to the quintet ($S = 2$) via intersystem crossing \cite{degraaf_study_2010,sousa_ultrafast_2013}.
Two types of triplet intermediate states, $^3$MLCT and metal-centered triplet $^3T_2$, have been observed experimentally on [Fe(bipy)$_3$]$^{2+}$ (bipy = 2,2$'$-bipyridine).
Ultrafast optical fluorescence measurements evidenced a $^3$MLCT state, while time-resolved spectroscopic measurements at the Fe $K$ edge identified a metal-centered intermediate state \cite{zhang_tracking_2014,kjaer_finding_2019}.
Time-resolved x-ray absorption of [Fe(phen)$_3$]$^{2+}$ (phen = 1,10-phenanthroline) at the Fe $M$ edge identified a metal-centered triplet $^3T_2$ state as intermediate state \cite{zhang_tracking_2019}.
Alías-Rodríguez \ea\ performed quantum wavepacket dynamics calculations on [Fe(bipy)$_3$]$^{2+}$, and suggest that the metal-centered triplet state is catalyzing the transfer to the high-spin state but also highlight the importance of bipyridine stretching vibrations to the photochemical pathway \cite{alias-rodriguez_ultrafast_2023}.
It therefore appears necessary to expand such investigation to complexes with other classes of ligands (to assess a possible influence of molecular vibrations on the relaxation pathway).

Quantum chemistry \cite{degraaf_study_2010,sousa_ultrafast_2013} and time-dependent density functional theory (DFT) \cite{dierks_ground-_2020} calculations have been successfully employed in the past to describe spin-crossover complexes.
Here, we use both static and time-dependent DFT, see \si, since this allows us to calculate potential energy surfaces of different electronic states including the ligand orbitals.
The atomic configuration in the intermediate $S = 1$ state is obtained by performing geometry optimization in a spin-restricted DFT calculation.
Subsequent linear-response time-dependent DFT calculations allow us to address electronic excitations not only in the Fe ion, but in the whole molecule.
The determined energies include all electrons with their interactions.
Fig.~\ref{fig:energySurface} shows the obtained potential energy surfaces of the ground state $S=0$ (blue), the singlet $^1$MLCT (turquoise area), $^3$MLCT (gray area), as well as the metal-centered triplet $S = 1$ (brown) and quintet $S = 2$ (red) as a function of the nuclear coordinate.
The latter is represented as the Fe-N distance change of one bond measured from the intermediate $S = 1$ state towards the $S = 0$ initial state (negative) and towards the $S = 2$ final state (positive values).

Based on these calculated potential energy surfaces we discuss possible reaction paths sketched by the arrows indicated in Fig.~\ref{fig:energySurface} to guide the reader.
The optical pump at 3.1 eV photon energy induces a resonant transition from the low-spin configuration to the $^1$MLCT manifold following the dipole selection rule, as determined experimentally on related systems \cite{bram_polychromatic_2012,aubock_femtosecond_2012}.
The transition to the final quintet state necessitates the evolution over an intermediate triplet state, which is provided by the ligand-based triplets and/or the metal-centered $S=1$ state.
Both scenarios lead to different possible relaxation pathways illustrated in Fig.~\ref{fig:energySurface}.
(i) Transfer from $^1$MLCT to ligand-based triplets, followed by transfer to the quintet potential energy surface (uppermost black curved arrow) and relaxation to the high-spin configuration.
(ii) Decay of the single $^1$MLCT state into multiple electronic excitations involving $S=1$ (vertical solid black arrow) followed by N-displacement to $S=2$ (horizontal arrow).
(iii) Relaxation at the low (or intermediate) spin nuclear configuration by subsequent steps (curved arrows) to vibrationally excited $S=1$ and N displacement to reach the high-spin state (horizontal arrow).
We note that for (ii) and (iii), the $^3$MLCT state may be populated prior to the metal-centered $S=1$ state.

With the likely contribution of the Fe-based triplet state, based on previous investigations \cite{zhang_manipulating_2017,kjaer_finding_2019,zhang_tracking_2019,alias-rodriguez_ultrafast_2023} and supported by the x-ray spectroscopy data, we propose that the population of the optically excited intermediate state, see Fig.~\ref{fig:timeEvolution}a (bottom), represents a nuclear wave packet evolving on the $S=1$ potential energy surface.
Pathways (ii-iii) agree with this proposal.
Our experimental data does not allow to distinguish between (ii) and (iii).
In addition, theoretical investigations \cite{penfold_spin-vibronic_2018} have shown the strong mixing of electronic, nuclear, and spin degrees of freedom suggesting more complex relaxation paths and both could contribute.
Furthermore, we cannot discard the possibility of having excited metal-centered triplet states at higher energies.

\begin{figure}[hbt]
	\includegraphics[width=\columnwidth]{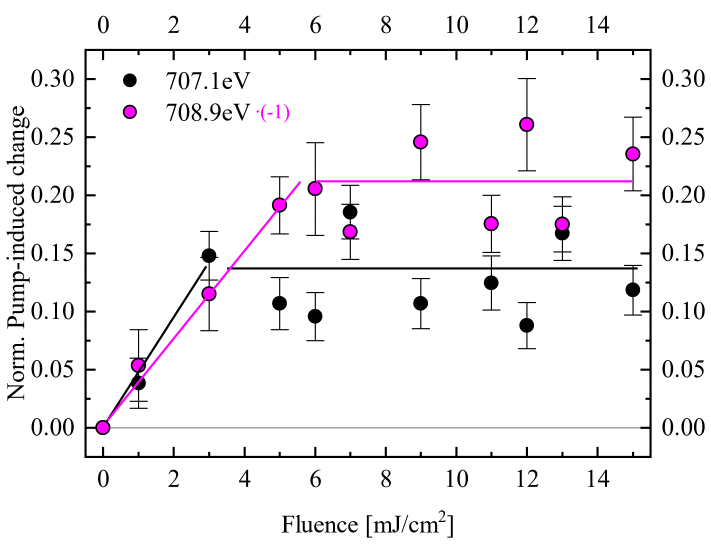}
	\caption{\footnotesize \textbf{Pump fluence dependence of the pump-induced change.}
The normalized pump-induced change at 3 ps time delay is plotted as a function of incident pump fluence in the x-ray absorption at the two features in the fine structure as indicated by the given photon energy.
For clarity, the absolute value of the negative pump-induced change at 708.9 eV photon energy is depicted.
The considerable fluctuations of the pump-induced changes are assigned to modifications in spatial overlap of pump and probe pulse upon varying the fluence.
Within the available data quality, we identify an increase in the pump-induced change to about 5 mJ/cm$^2$ and a saturation of these signals for larger pump fluence.
	\label{fig:fluence}}
\end{figure}

An essential difference between molecular films and individual molecules in solution is the presence of molecule-molecule interactions in the condensed phase.
These interactions have an impact on the switching of the films, which may in turn be useful for applications \cite{cobo_multilayer_2006}.
Since the pump absorption determines the number of optically excited molecules, analysis of the switching yield as a function of pump fluence may provide insight into the interaction of excited molecules among each other in the molecular film.
Fig.~\ref{fig:fluence} shows the normalized pump-induced change as a function of incident pump fluence up to 15 mJ/cm$^2$ at a fixed pump-probe delay of 3 ps.
The absolute values of the pump-induced changes of the corresponding fine structure x-ray energies, which were introduced in Fig.~\ref{fig:scs}c, increase linearly with the fluence up to approximately 5 mJ/cm$^2$.
This behavior suggests that the excited molecules in this regime are too dilute for interaction among them to play a role.
For fluences of 3 to 6 mJ/cm$^2$ the changes saturate for both x-ray energies.
As detailed in the \si, a fluence of 13 mJ/cm$^2$ corresponds, in the investigated films, to approximately one absorbed pump photon per spin crossover molecule.
We estimate from the fluence at which the saturation is reached in Fig.~\ref{fig:fluence} that 30 to 50\,\% of the molecules in the film are transiently excited to the high-spin state.

Light-induced back switching to the low-spin state is expected to be negligible for 400\,nm light \cite{hauser_light-induced_2004}.
This implies that although sufficient pump photons reach the film, a large fraction of the molecules does not participate in the switching process.
Such a behavior was reported earlier for static measurements at low temperatures for the present molecule \cite{grunwald_defying_2023} as well as other complexes \cite{bairagi_molecular-scale_2016,johannsen_electron-induced_2021}, and rationalized by steric-repulsion considerations.
In other words, the space around some of the molecules is too small to accommodate a high-spin state.
For a fraction of the molecules, the high-spin state is unstable due to the interaction with neighboring molecules.
Such intermolecular interactions, modeled as elastic interactions between the molecules, were shown to cause incomplete spin transition in films \cite{kelai_thermal_2021} and one-dimensional chains \cite{weber_quenching_2008,chiruta_analysis_2014,traiche_elastic_2018}.
The HS fraction reached in the transient measurements is $\approx 10$\,\% lower than that of static ones at low temperatures.
The main cause of the incomplete switching in the static and ultrafast measurements is most likely the same, \ie\ steric repulsion.
The 10\,\% difference may be ascribed to stimulated emission under high fluence excitation.

\subsection*{Conclusions}
We investigated the photo-induced ultrafast spin-state switching dynamics of an \textit{ultrathin film} of Fe(II) complexes using transient x-ray absorption spectroscopy at the Fe $L_3$ edge.
Those measurements at the Fe $L_3$ edge, directly sensitive to the electronic occupation of the Fe $d$ orbitals evolving with the switching, were made possible by recent developments at modern light sources like x-ray free electron lasers along with the increased quality and robustness of the molecular thin films.
The switching from the low-spin to the high-spin state is completed within 500\,fs.
The transient evolution of the Fe $L_3$ fine structure evidences the implication of an intermediate state attributed to a metal-centered triplet state with the geometry, and thus with the ligand field, of the low-spin state.
We thereby confirm the recent observations in the photo-induced spin-state switching of solvated molecules\cite{zhang_tracking_2014} and thick films \cite{zhang_tracking_2019}.
Our results show, in addition, a saturation of the fraction of switched molecules to $\approx 50$\,\% with increasing optical-excitation fluence, which is likely caused by molecule-molecule interactions.
We highlight the sensitivity of this spectroscopy to the spin character of the electronically excited state correlated with the nuclear coordinate.
Considering different complexes, the relative importance of the $^3T_2$ and $^3$MLCT states is expected to depend on the actual strength of the ligand field \cite{zhang_manipulating_2017}.
In the future, the opportunities arising from our work to exploit the connection between intramolecular parameters, such as the ligand field, and structural parameters of intermolecular arrangement in condensed films will contribute to identifying microscopic interaction principles in these films.
Hence, combining structure-specific interactions resulting from the molecular arrangement in thin films with intra-molecular, chemically controlled characteristics, such as the ligand field strength, may open up further opportunities for material design.

\subsection*{Methods}
Details on sample preparation and characterization, data acquisition, and calculations are provided in the \si.

\begin{acknowledgement}
\footnotesize
We acknowledge European XFEL in Schenefeld, Germany, for provision of x-ray free electron laser beamtime at Scientific Instrument Spectroscopy and Coherent Scattering and would like to thank their staff for the excellent assistance before, during, and after the beamtime.
We thank Thies J. Albert, Jennifer Schmeink, Ulrich von Hörsten, Samuel Palato, Sergey Kovalenko, and Julia Stähler for supporting measurements as well as Christel Marian, Eberhard K. U. Gross and Georg Jansen for fruitful discussions.
This work was funded by the Deutsche Forschungsgemeinschaft (DFG, German Research Foundation) Project No. 278162697-SFB 1242.
G. K. and P.K. gratefully acknowledge the computing time granted by the Center for Computational Sciences and Simulation (CCSS) of the University of Duisburg-Essen and provided on the supercomputer magnitUDE (DFG grants No. INST 20876/209-1FUGG and No. INST 20876/243-1 FUGG) at the Zentrum für Informations und Mediendienste (ZIM).
F.T. acknowledges funding from the German Research Foundation with the grant TU58/18-1.
\end{acknowledgement}

\subsection*{Data availability}
Data recorded for the experiment at the European XFEL are available at DOI: 10.22003/XFEL.EU-DATA-002783-00.

\subsection*{\si}
\si\ on the sample preparation, time-resolved x-ray absorption spectroscopy measurement setup, determination of time-resolved x-ray absorption spectra, multiplet calculations, pumped XAS spectra in the femtosecond time regime, relaxation to the low-spin ground state, fragmentation of molecules under illumination with large fluence, optical microscopy, homogeneity of the sample, UV-Vis spectroscopy, density functional theory, and the three state model.

\providecommand{\latin}[1]{#1}
\makeatletter
\providecommand{\doi}
  {\begingroup\let\do\@makeother\dospecials
  \catcode`\{=1 \catcode`\}=2 \doi@aux}
\providecommand{\doi@aux}[1]{\endgroup\texttt{#1}}
\makeatother
\providecommand*\mcitethebibliography{\thebibliography}
\csname @ifundefined\endcsname{endmcitethebibliography}
  {\let\endmcitethebibliography\endthebibliography}{}

\end{document}